\documentclass[conference]{IEEEtran}
\IEEEoverridecommandlockouts
\usepackage{cite}
\usepackage{amsmath,amssymb,amsfonts}
\usepackage{algorithmic}
\usepackage{graphicx}
\usepackage{textcomp}
\usepackage{xcolor}
\usepackage{util}
\usepackage{multicol}
\usepackage{multirow}
\usepackage{hyperref}
\usepackage{float}
\usepackage{amsmath}
\usepackage{amssymb}
\usepackage{mathtools}
\usepackage{pgfplots}
\pgfplotsset{compat=1.18}
\usepackage{amsthm}

\def\BibTeX{{\rm B\kern-.05em{\sc i\kern-.025em b}\kern-.08em
    T\kern-.1667em\lower.7ex\hbox{E}\kern-.125emX}}
\begin{document}

\title{Goal-guided Generative Prompt Injection Attack on Large Language Models\\
}

\author{
    \IEEEauthorblockN{Chong Zhang\textsuperscript{1}, 
                      Mingyu Jin\textsuperscript{1}, 
                      Qinkai Yu\textsuperscript{2}, 
                      Chengzhi Liu\textsuperscript{1}, 
                      Haochen Xue\textsuperscript{1}, 
                      Xiaobo Jin\textsuperscript{1,$^{\dagger}$} \thanks{$^{\dagger}$ Corresponding Author. 
                        This work was partially supported by Research Development Fund with No. RDF-22-01-020, the “Qing Lan Project” in Jiangsu universities and National Natural Science Foundation of China under Grant U1804159.}}
    \IEEEauthorblockA{\textsuperscript{1}Xi'an Jiaotong-Liverpool University, 
                      \textsuperscript{2}University of Liverpool\\
                      \textbf{Email:} Chong.zhang19@student.xjtlu.edu.cn, Xiaobo.jin@xjtlu.edu.cn}
}

\maketitle

\begin{abstract}

Current large language models (LLMs) provide a strong foundation for large-scale user-oriented natural language tasks. Numerous users can easily inject adversarial text or instructions through the user interface, thus causing LLM model security challenges. Although there is currently much research on prompt injection attacks, most of these black-box attacks use heuristic strategies. It is unclear how these heuristic strategies relate to the success rate of attacks and thus effectively improve model robustness. To solve this problem, we redefine the goal of the attack: to maximize the KL-divergence between the conditional probabilities of the clean text and the adversarial text. Furthermore, we prove that maximizing the KL-divergence is equivalent to maximizing the Mahalanobis distance between the embedded representation $x$ and $x'$ of the clean text and the adversarial text when the conditional probability is a Gaussian distribution and gives a quantitative relationship on $x$ and $x'$. Then we designed a simple and effective goal-guided generative prompt injection strategy (G2PIA) to find an injection text that satisfies specific constraints to achieve the optimal attack effect approximately. It is noteworthy that our attack method is a query-free black-box attack method with a low computational cost. Experimental results on seven LLM models and four datasets show the effectiveness of our attack method.

\end{abstract}


\section{Introduction}

Large Language Models (LLMs) \cite{openai2023gpt4} are evolving rapidly in terms of architecture and applications. As they become more and more deeply integrated into our lives, the urgency of reviewing their security properties increases. A large number of previous studies \cite{kaufmann2023survey} have shown that LLMs whose instructions are adjusted through reinforcement learning with human feedback (RLHF) are highly vulnerable to adversarial attacks. Therefore, studying adversarial attacks on large language models is of great significance, which can help researchers understand the security and robustness of large language models \cite{shayegani2023survey} and thus design more powerful and robust models to prevent such attacks.

Attacks against large models are mainly divided into white-box attacks and black-box attacks. White-box attacks assume that the attacker has full access to the model weights, architecture, and training process to obtain gradient information, so the main method is gradient-based attacks. Guo et al. \cite{guo2021gradientbased} proposed a new gradient-based distributed attack (GBDA), which uses the Gumbel-Softmax approximation technique to make the adversarial loss optimization differentiable on the one hand, and uses BERTScore and perplexity to enhance perceptibility and fluency on the other hand. Other white-box methods like DeepFool \cite{moosavi2016deepfool} and TextBugger \cite{li2018textbugger} rely on calculating the model gradient to attack the model. TextBugger uses the gradient information to find the most effective text perturbation to complete the attack. DeepFool determines the minimum perturbation to find the boundary of sample classification error to achieve the attack goal. The disadvantage of these methods is that they can only attack large open-source language models. Still, for more widely used closed-source LLMs such as ChatGPT, these methods are powerless because they cannot obtain the architecture and parameters of the model.

Compared with white-box attacks, black-box attacks limit attackers to access only API-type services. According to the attack granularity, black-box attacks \cite{goyal2023survey} can be divided into letter level, word level, sentence level, and multi-level. Most black-box attacks use word replacement \cite{jin2020bert}: just find the most critical words in the text and replace them, or use some simple text enhancement methods \cite{morris2020textattack}, such as synonym replacement, random insertion, random exchange, or random deletion. Since black-box strategies cannot know the internal structure of large models, most attack methods use heuristic strategies. However, it is not clear how these heuristic strategies are related to the success rate of attacks, and how to effectively improve the success rate of attacks.

In our work, we assume that the clean text representation $x$ and the adversarial text representation $x'$ satisfy the conditional probability distribution $p(y|x)$ and $p(y|x')$ respectively, and the goal of the black-box attack is to maximize the KL-divergence $\tm{KL}(p(y|x),p(y|x'))$, then we prove that maximizing the KL-divergence is equivalent to maximizing the Mahalanobis distance between $x$ and $x'$ under the assumption of Gaussian distribution. Furthermore, we give the quantitative relationship between optimal attack text representation ${x'}^{*}$ and $x$. Based on the above theoretical results, we designed a simple and effective prompt text injection method to search for attack texts that meet approximately optimal conditions. 

Overall, our contributions are as follows: \tf{1)} We propose a new objective function based on KL-divergence between two conditional probabilities for black-box attacks to maximize the success rate of black-box attacks; \tf{2)} We theoretically prove that under the assumption that the conditional probabilities are Gaussian distributions, the KL-divergence maximization problem based on the posterior probability distributions of clean text and adversarial text, respectively, is equivalent to maximizing the Mahalanobis distance between clean text and adversarial text. \tf{3)}We propose a simple and effective injection attack strategy for generating adversarial injection texts, and the experimental results verify the effectiveness of our method. Note that our attack method is a query-free black-box attack method with low computational cost.

\section{Related Work}

In this section, we mainly review white-box and black-box attacks. Gradient-based attacks are in the white-box category, while token manipulation and prompt injection are all black-box attacks.

\subsection{Gradient-based Attack}  Gradient-based distribution attack (GBDA) \cite{guo2021gradientbased} used Gumbel-Softmax approximation to make adversarial loss optimization differentiable on the one hand, and also uses BERTScore and perplexity to enhance perceptibility and fluency on the other hand. HotFlip \cite{ebrahimi2018hotflip} mapped text operations to vector space and calculated the derivative of the loss based on these vectors. Based on different tasks, AutoPrompt \cite{shin2020autoprompt} used a gradient-based search strategy to find the most suitable prompt template for them. Autoregressive random coordinate ascent (ARCA) \cite{jones2023automatically} searched for input-output pairs that match specific behavioral patterns under the broader optimization problem framework. Wallace et al. \cite{wallace2021universal} proposed a gradient-guided token search method for searching for 5 tokens for short sequences, 1 for classification and 4 for generation, called universal adversarial triggers (UAT), which are used to trigger the model to produce specific predictions. However, gradient-based white-box attacks are not suitable for widely used closed-source large language models.


\subsection{Token Manipulation Attack} Given a text input containing a sequence of tokens, simple word manipulations (e.g. replacing them with synonyms) can trigger the model to make incorrect predictions. Ribeiro et al. \cite{ribeiro2018semantically} manually defined adversarial rules (SEAR) that are semantically equivalent to heuristics to perform minimal token manipulations, thereby improving the attack success rate of the model. In contrast, Easy Data Augmentation (EDA) \cite{wei2019eda} defined a set of simpler and more general operations to augment the text, such as synonym replacement, random insertion, random swapping, or random deletion. BERT-Attack \cite{li2020bert} replaces words with semantically similar ones through BERT based on the fact that context-aware prediction is a natural use case for masked language models. Unlike the above attack methods, our attack method attempts to insert generated adversarial prompts instead of just modifying certain words in the prompt.

\subsection{Prompt Injection Attack}  Generally, larger LLMs have excellent instruction tracking capabilities and are more vulnerable to attacks by malicious actors. An attacker \cite{mckenzie2023inverse} can easily embed instructions in the data and trick the model into misinterpreting it. In the work of Perez\&Ribeiro \cite{perez2022ignore}, the goals of prompt injection attacks are divided into target hijacking and prompt leakage. The former attempts to redirect the original target of the LLM to a new target desired by the attacker, while the latter attempts to persuade the LLM to disclose the initial system prompt of the model program. Obviously, system prompts are very valuable to companies because they can significantly affect model behavior and change user experience. Liu et al. \cite{liu2023prompt} found that LLMs are highly sensitive to escape characters and delimiters appearing in prompts, which implicitly convey instructions to start a new scope within the prompt by escaping. It is worth noting that our method for generating prompt injection attacks does not attempt to insert manually specified attack instructions, but instead attempts to influence the output of LLMs by generating confusing prompts based on the original prompts.

\subsection{Black-box Attack Paradigm}
The black box attack paradigm refers to a broad range of attack methods that exploit vulnerabilities in machine learning models. Boundary query-free attack BadNets \cite{gu2017badnets} and boundary query-related attack \cite{goodfellow2014explaining} are two common black box attack methods, whose difference lies in whether the boundaries of the model need to be input in advance to observe the output of the model. Other black box attack methods include methods based on model substitution, such as practical black box attack\cite{papernot2017practical} and context injection-based attacks, BERT-Attack \cite{li2020bert}.


\section{Methodology}

\subsection{Threat Model with Black-box Attack}

\subsubsection{Adversarial scope and goal.} Given a text $t$ containing multiple sentences, we generate a text $t'$ to attack the language model, ensuring that the meaning of the original text $t$ is preserved, otherwise, then we believe that the attack text $t'$ is attacking another text that is unrelated to $t$. Here, we use $\mc{D}(t',t)$ to represent the distance between the semantics of text $t$ and $t'$. If the LLM outputs $M(t)$ and $M(t')$ differ, then $t'$ is identified as an adversarial input for $M$. Our objective is formulated as follows:
\eqn{\label{eqn:LLM-attack}}{
    M(t)=r, \quad M\left(t^{\prime}\right)=r^{\prime}, \quad \mc{D}(r,r^{\prime}) \ge \varepsilon, \quad \mathcal{D}\left(t^{\prime}, t\right) < \varepsilon,
}
where the texts $r$ and $r'$ are the outputs of model $M$ on text $t$ and $t'$ respectively, and $r$ is also the groundtruth of text $t$. We introduce a distance function $\mc{D}(\cdot,\cdot)$ and a small threshold $\varepsilon$ to measure the semantic relationship between two texts. In our problem, we aim to provide the following attack characteristics

\im{
\item \textbf{Effective:} Problem \ref{eqn:LLM-attack} shows that the attack model ensures a high attack success rate (ASR) with $\mc{D}(M(t'),r) \ge \varepsilon$ on one hand and maintains high benign accuracy with $\mc{D}(M(t),r) < \varepsilon$ on the other hand.

\item \textbf{Imperceptible:} Prompt injection attacks are often detected easily by inserting garbled code that can disrupt large models. We try to ensure that the adversarial text fits better into the problem context so that the model's active defense mechanisms make it difficult to detect the presence of our prompt injections. 

\item \textbf{Input-dependent:} Compared to fixed triggers, input-dependent triggers are imperceptible and difficult to detect from humans in most cases \cite{wallace2022benchmarking}. It is obvious the adversarial text (or trigger) $t'$ is input-dependent by Eqn. (\ref{eqn:LLM-attack}). We will insert trigger $t'$ into $t$ through a prompt injection to form an attack prompt (see Sec. \ref{sn:prompt-injection}).
}

\subsection{Analysis on Objective}
For the convenience of discussion, we regard the text generation of LLM as a classification problem, where the output will be selected from thousands of texts and each text is regarded as a category. Below, we first discuss the necessary conditions for the LLM model to output different values ($M(t) \neq M(t')$) under the conditions of clean text $t$ and adversarial text $t'$, respectively.

As can be seen from the problem (\ref{eqn:LLM-attack}), the input $t$ and output $r$ of model $M$ are both texts. To facilitate analysis, we discuss the embedded representation of texts in the embedding space.  Assume that two different text inputs, $t$ and $t'$, are represented by vectors $x = w(t)$ and $x' = w(t')$, respectively. The corresponding outputs $r$ and $r'$ are represented by vectors $y = w(r)$ and $y' = w(r')$, where $w(\cdot)$ is a bijective function between text and vector. Since we assume that the outputs $r$ and $r'$ of LLM come from an enumerable discrete space and there is a one-to-one correspondence between text representations and vector representations, their vector representations $y$ and $y'$ are also enumerable. Now, we re-formulate the output of the LLM model $M$ as a maximization problem of posterior probability in an enumerable discrete space $\mc{Y}$
\begin{equation}
\begin{split}
y &= \argmax_{\hat{y} \in \mathcal{Y}} p(\hat{y}|x) = w(M(w^{-1}(x))), \\
y' &= \argmax_{\hat{y} \in \mathcal{Y}} p(\hat{y}|x') = w(M(w^{-1}(x'))),
\end{split}
\end{equation}
where $w^{-1}(\cdot)$ is the inverse function of $w(\cdot)$ the function. Obviously, we have
\eqn{}{
\forall \hat{y},\quad p(\hat{y}|x) = p(\hat{y}|x') \Rightarrow \argmax_{\hat{y} \in \mc{Y}} p(\hat{y}|x) = \argmax_{\hat{y} \in \mc{Y} } p(\hat{y}|x').
}
So, we get its converse proposition
\eqn{}{
\argmax_{\hat{y} \in \mc{Y}} p(\hat{y}|x) \neq \argmax_{\hat{y} \in \mc{Y}} p(\hat{y}|x') \Rightarrow \exists \hat{y},\quad p(\hat{y}|x) \neq p(\hat{y}|x').
}
That is, LLM has different posterior probability distributions under different input conditions, which is necessary for LLM to output different values such as $M(t) \neq M(t')$. To increase the likelihood that LLM will output different values, we quantify the divergence between the probability distributions $p(y|x)$ and $p(y|x')$ with Kullback-Leibler (KL) divergence and maximize it
\begin{equation}
    \max_{x'} \mathrm{KL}(p(y|x), p(y|x')).
    \label{eqn:kl-surrogate}
\end{equation}

\begin{figure}[!htp]
\centering
\vspace{-5pt}
\begin{tikzpicture}
    \draw plot coordinates {(-0.5, 0.008863696823876015) (-0.4, 0.015830903165959937) (-0.30000000000000004, 0.02716593846737123) (-0.20000000000000007, 0.044789060589685764) (-0.10000000000000009, 0.07094918569246285) (-1.1102230246251565e-16, 0.10798193302637613) (0.09999999999999987, 0.15790031660178824) (0.19999999999999984, 0.22184166935891106) (0.2999999999999998, 0.2994549312714896) (0.3999999999999998, 0.38837210996642574) (0.4999999999999998, 0.48394144903828656) (0.5999999999999996, 0.5793831055229652) (0.6999999999999997, 0.6664492057835991) (0.7999999999999998, 0.7365402806066466) (0.8999999999999997, 0.7820853879509118) (0.9999999999999996, 0.7978845608028654) (1.0999999999999996, 0.782085387950912) (1.1999999999999997, 0.7365402806066469) (1.2999999999999996, 0.6664492057835997) (1.3999999999999995, 0.579383105522966) (1.4999999999999996, 0.4839414490382872) (1.5999999999999996, 0.38837210996642624) (1.6999999999999993, 0.29945493127149037) (1.7999999999999994, 0.2218416693589116) (1.8999999999999995, 0.1579003166017886) (1.9999999999999996, 0.1079819330263763) (2.0999999999999996, 0.07094918569246297) (2.1999999999999993, 0.04478906058968594) (2.2999999999999994, 0.027165938467371323) (2.3999999999999995, 0.01583090316595998) };
    \draw[->] (-1.0, 0) to (3.0, 0);
    \draw[->] (0, -0.5) to (0, 1.5);
    \node[above] at (1, 0) { $x$ };
    \fill (1, 0) circle (0.05cm);
    \node[above] at (1, 0.7978845608028654) { $p(y|x)$ };
    \draw[dashed] (0.75, 0) to (0.75, 0.704130653528599);
    \node[below] at (0.75, 0) { $y$ };
    \draw[dashed] (1.4, 0) to (1.4, 0.5793831055229657);
    \node[below] at (1.4, 0) { $y_1$ };
    \draw[dashed] (0.2, 0) to (0.2, 0.2218416693589111);
    \node[below] at (0.2, 0) { $y_2$ };
\end{tikzpicture}
\vspace{-5pt}
\caption{Assumption that the output $y$ of LLM under the condition of $x$ satisfies the discrete Gaussian distribution:
Answers (output) $y$ close to question (input) $x$ are usually more relevant to $x$ and have a higher probability of being sampled.}
\vspace{-7pt}
\label{fig:discrete-gaussian}
\end{figure}
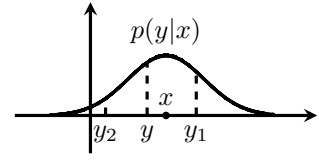

First, we assume that the output distribution $p(y|x)$ of LLM satisfies the discrete Gaussian distribution \cite{canonne2021discrete} under the condition of input $x$, 
\eqn{}{
p(y|x) = \frac{e^{- \frac{1}{2}(y - x)^T \Sigma^{-1}(y - x)} }{
\sum_{\hat{y} \in \mc{Y} }e^{- \frac{1}{2}(\hat{y} - x)^T \Sigma^{-1}(\hat{y} - x)}
},
}
as shown in Fig. \ref{fig:discrete-gaussian}, that is, the output of LLM is defined on a limited candidate set $\mc{Y}$, although this candidate set $\mc{Y}$ may be very large. Since the input $x$ and output $y$ of LLM are questions and answers, the spaces where they are located generally do not intersect with each other, so there are $y = \argmax_{\hat{y} \in \mc{Y}} p(\hat{y}|x) \neq x$,
which is different from the commonly used continuous Gaussian distribution $y = x$, as can be seen in Fig. 1.  For the same question $x$, LLM usually outputs different answers and the answer $y$ most relevant to $x$ has a higher probability of being sampled. In the embedding space, the distance between $y$ and $x$ is usually closer. Similarly, answers $y$ that are almost uncorrelated with $x$ and far away from $x$ in the embedding space have a smaller probability of being sampled.

To solve the problem (\ref{eqn:kl-surrogate}) theoretically, we do the following processing: Replace the discrete Gaussian distribution with a continuous Gaussian distribution to facilitate the calculation of KL-divergence although the output text on condition of input text still obeys discrete distribution in practical applications. Subsequently, we present the following theorem (see Appendix Section A and B for details).

\tf{Theorem 1} Assuming $p(y|x)$ and $p(y|x')$ respectively follows the Gaussian distribution $\mc{N}_1(y;x,\Sigma)$ and $\mc{N}_2(y;x',\Sigma)$, then the maximization $KL(p(y|x),p(y|x'))$ is equivalent to maximizing the \tf{Mahalanobis distance} $(x' - x)^T \Sigma^{-1}(x' - x)$, which is further transformed into the following minimization optimization problem given the clean input $x$
\eqn{\label{eqn:opt-problem-text}}{
   \min_{x'} \|x'\|_2 , \quad s.t. \quad (x' - x)^T \Sigma^{-1}(x' - x) \le 1,
}
which has an optimal solution of the form ($\lambda$ is Lagrange multiplier)
\eqn{}{
   x'^* = (\Sigma + \lambda I)^{-1} \lambda x,\quad \lambda > 0.
}

\ssn{Solving Problem (\ref{eqn:opt-problem-text}) Approximately via Cos Similarity}
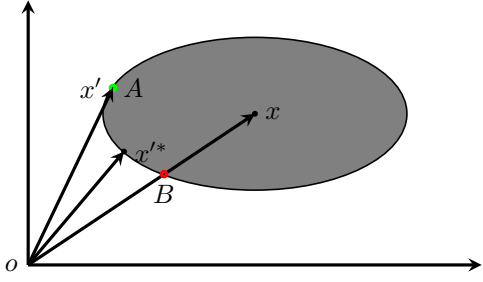
\begin{figure}[!htp]
\centering
\begin{tikzpicture}
    \coordinate (x) at (3,2);
    \draw (x) ellipse (2 and 1);
    \fill[gray] (x) ellipse (2 and 1);
    \draw[->] (0,0) -- (6,0);
    \draw[->] (0,0) -- (0,3.5);
    
    \draw (x) circle (0.5pt);
    \node[right] at (x) {$x$};
    
    \coordinate (z) at ({3 + 2*cos(210)},{2 + 1*sin(210)});
    \draw (z) circle (0.5pt);
    \node[right] at (z) {$x'^{*}$};
    \draw[->] (0,0) -- (z);
    \draw[->] (0,0) -- (x);

    \coordinate (A) at ({3 + 2*cos(160)},{2 + 1*sin(160)});
    \draw[color=green] (A) circle (1pt);
    \draw[->] (0,0) -- (A);
    
    \node[right] at (A) {$A$};
    \node[left] at (A) {$x'$};

    \coordinate (B) at ({3 + 2*cos(233)},{2 + 1*sin(233)});
    \node[below] at (B) {$B$};
    \draw[color=red] (B) circle (1pt);
    
    \node[left] at (0,0) {$o$};
\end{tikzpicture}
\vspace{-5pt}
\caption{Assuming $x'^* = (x'_1,x'_2)$ is the optimal solution to problem (\ref{eqn:opt-problem-text}), then when $x'$ moves from $A$ through $x'^*$ to $B$ on the ellipse, $\cos(x',x)$ first increases and then decreases, while $\|x'\|_2$ first decreases and then increases.}
\vspace{-5pt}
\label{fig:ellipse}
\end{figure}

Note that our method is a black-box attack and does not know the model parameters $\Sigma$, so we cannot solve the problem (\ref{eqn:opt-problem-text}). Below, we try to approximately solve the problem (\ref{eqn:opt-problem-text}) using cos similarity, which does not contain any parameters. Fig. \ref{fig:ellipse} shows the optimal solution $x'^*$ of the problem (\ref{eqn:opt-problem-text}) in two-dimensional space. When the vector $x'$ moves from $A$ to $B$ along the ellipse through the optimal point $x'^*$, $\cos(x',x)$ first increases and then decreases, while $\|x'\|_2$ first decreases and then increases. Therefore, we introduce the hyperparameter $\gamma$  to approximate the solution to problem (\ref{eqn:opt-problem-text})
\eqn{}{
\cos(x',x) = \gamma,\quad 0 \le \gamma \le 1,
}
where $x$ (known) and $x'$ (unknown) are the embedded representations of clean and adversarial input, respectively. Note that in our implementation, we relax the constraint satisfaction problem of the optimal solution $x'^*$ as
\eqn{}{
|\cos(x',x) - \gamma| < \delta,\quad \delta \tm{ is a small positive constant.}
}

Below we discuss the problems of $\gamma \neq 0$ and $\gamma \neq 1$ from two perspectives. First, we prove this conclusion mathematically. We compute $\cos(x'^*,x)$ to obtain
\eqn{}{
\cos(x'^*,x) = \frac{{x'^*}^{T} x}{\|x'^*\|_2\|x\|_2} = \frac{\lambda x^T (\Sigma + \lambda I)^{-1} x}{\|x'^*\|_2\|x\|_2}.
}
If $\cos(x'^*,x) = 0$ ($x \neq 0$), then $\lambda = 0$, that is, $x'^* = 0$, which is meaningless to LLM. If $\cos(x'^*,x) = 1$, then $x'^*$ and $x$ are in the same direction, i.e. $x'^* = \lambda (\Sigma + \lambda I)^{-1}x = t x$, where $t$ is a ratio value. So $x'^*$ must be the eigenvector of the matrix $\lambda (\Sigma + \lambda I)^{-1}$. However, $x'^*$ can be an embedding representation of any input. So we arrive at a contradiction. 

From the perspective of a black-box attack, when $\cos(x',x) = 1$, the vectors $x$ and $x'$ will have the same direction.  When using vectors (often using unit vectors) to represent text, we care more about the direction of the vector, so $x = x'$. In addition, note that $w(\cdot)$ is a bijective function, then for clean text $t$ and adversarial text $t'$, there is $t = t'$. There we have $r = M(t) = M(t') = r'$, which contradicts the condition $r \neq r'$ in problem (\ref{eqn:LLM-attack}). When $\cos(x',x) = 0$, then $w(t')$ and $w(t)$ are linearly uncorrelated, which is conflicted with the condition $D(t',t) = 0$ in problem (\ref{eqn:LLM-attack}).
\subsection{Goal-guided Generative Prompt Injection Attack} \label{sn:prompt-injection}
\begin{figure*}[!ht]
    \centering
    \vspace{-20pt}
    \includegraphics[width=0.7\textwidth]{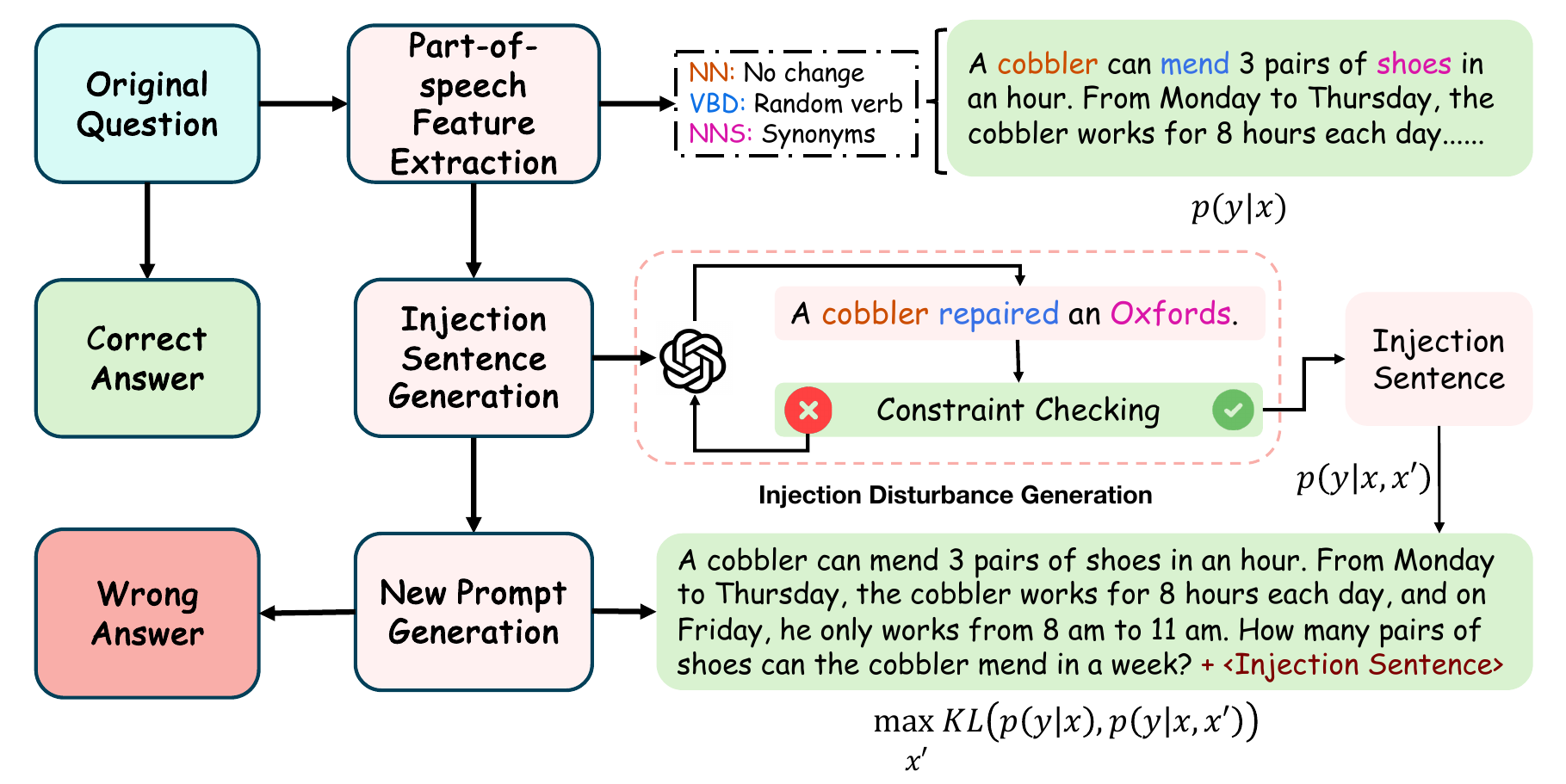}
    \vspace{-10pt}
    \caption{Overview of Goal-guided Generative Prompt Injection Attack:  1) We use the part-of-speech method to find the subject, predicate and object of the question in the clean text $x$ and fetch synonyms of the predicate and object plus a random number as core words; 2) Put the core words into assistant LLM to generate an adversarial text $x'$ that satisfies the constraints; 3) Insert the generated adversarial text into the clean text $x$ to form the final attack text; 4) Enter the attack text into the LLM victimization model to test the effectiveness of our attack strategy.}
    \vspace{-10pt}
    \label{fig:overview}
\end{figure*}

Note that $w(t) = x$ and $w(t') = x'$. Based on the previous discussion, we can simplify our problem (\ref{eqn:LLM-attack}) into the following constraint satisfaction problem (CSP)
\eqna{
\min_{t'} && 1, \\
s.t. && \mc{D}(t',t) < \epsilon, \label{eqn:dist-condition} \\
&& |\cos(w(t'),w(t)) - \gamma| < \delta, \label{eqn:cos-condition}
}
where $x$ and $x'$ represent clean input (known) and adversarial input (unknown), respectively, while $w(\cdot)$ represents the embedded representation of the text (literal meaning) and $\mc{D}(\cdot,\cdot)$ represents the distance between the semantics (intrinsic meaning) of the two texts. The hyperparameters $\delta$ and $\epsilon$ are used to control the difficulty of searching constraint, where $\delta$ or $\epsilon$ is smaller, the search accuracy is higher. 

In our method, we implement a black-box attack through prompt injection: generate an adversarial text $t'$ that satisfies conditions (\ref{eqn:cos-condition}) and (\ref{eqn:dist-condition}), and then mix $t'$ into the text $t$ to obtain a prompt $\bar{t}$.  The advantage of using prompt insertion is that since the prompt $\bar{t}$ contains both clean input $t$ and adversarial input $t'$, on one hand, the concealment of the adversarial input (or trigger) $t'$ is enhanced, and on the other hand the adversarial input $t'$ plays a very good interference role to the output of the LLM model.

Next, we first find the core word set that determines the text semantics through the semantic constraint condition (\ref{eqn:dist-condition}) and then use the core word set to generate adversarial text that satisfies the cos similarity condition (\ref{eqn:cos-condition}). It is worth noting that the embedded representation is defined on \tf{texts}, so we use the\tf{BERT}model to convert any text $t$ into $w(t)$. However, in our work, the semantic distance between texts is defined on the core \tf{words} of the text, so we use the \tf{word2vec} model to define the semantic distance $\mc{D}(t',t)$ between text $t'$ and $t$.

\subsection{Solving Semantic Constraint(\ref{eqn:dist-condition})}
Usually, the semantic meaning of text $r$ is determined by a few core words $C(r) = \{\omega_1(r),\omega_2(r),\cdots,\omega_n(r)\}$, which will not be interfered by noise words in the text $r$. Based on the core word set $C(r)$, we will use cos similarity to define the semantic distance $\mc{D}(t',t)$ between two texts $t$ and $t'$
\eqn{\label{eqn:word-constraint}}{
    \mc{D}(t',t) = 1 - \cos(s(\omega_i(t')),s(\omega_i(t))),\quad i = 1,2,\cdots,n,
}
where $s(\cdot)$ represents the word2vec representation of the word.

In a text paragraph, usually, the first sentence is a summary of the entire paragraph (maybe some exceptions that we ignore), where the meaning of a text will be determined by its subject, predicate, and object. Therefore, the subject $S_t$, predicate $P_t$ and object $O_t$ that appear first in text $t$ will serve as the core words of the text ($n = 3$)
\eqn{}{
\omega_1(t) = S_t,\quad \omega_2(t) = P_t, \quad \omega_3(t) = O_t.
}
Because the change of the subject will have a great impact on the meaning of the text, the subject $S_{t'}$ in the adversarial text $t'$ is directly set to the subject $S_t$ in the clean text $t$. Through the WordNet tool, we randomly select a word from the synonym lists of $P_t$ or $O_t$ to check whether the constraints (\ref{eqn:dist-condition}) are met. Once the conditions are met, the search process will stop. Finally, we obtain the core word set of the adversarial text $t'$ that satisfies the semantic constraints
\eqn{}{
C(t') = \{\omega_1(t') = S_{t'} = S_t,\quad \omega_2(t') = P_{t'}, \quad \omega_3(t') = O_{t'} \}.
}

\ssn{Solving Cos Similarity Constraint (\ref{eqn:cos-condition})}
Next, we will generate adversarial text that satisfies constraint (\ref{eqn:cos-condition}) through the core vocabulary $C(t')$ of adversarial text. Note that to increase the randomness of the sentence, we add another random number $N_{t'}$ between $10$ and $100$ as the core word. Now the core vocabulary of the adversarial text $t'$ becomes ($n = 4$)
\begin{equation}
\begin{split}
C(t') = \{ & \omega_1(t') = S_{t'}, \omega_2(t') = P_{t'}, \\
           & \omega_3(t') = O_{t'}, \omega_4(t') = N_{t'} \}.
\end{split}
\end{equation}

The core word set is embedded into the prompt template to generate a sentence text $t^\prime$ with LLM.  We iterate $N$ times to randomly generate multiple sentence texts $t^\prime$ until the text $t^\prime$ satisfies Eqn. (\ref{eqn:cos-condition}). Finally, we insert the adversarial text $t^\prime$ after the text $t$ to attack the LLM. Inserting $t^\prime$ after any sentence in $t$ is also feasible. In Appendix Section G, we will see minimal difference in attack effectiveness at different locations.

\section{Experiments}

\subsection{Experimental Details}

Below we describe some details of the prompt insertion-based attack method, including the victim model, dataset, and evaluation metrics. In particular, ChatGPT-4-Turbo (gpt-4-0125-preview) is used as our auxiliary model to generate random sentences that comply with grammatical rules. Unless otherwise stated, all results of our algorithm use the parameter settings $\epsilon = 0.2$, $\delta = 0.05$ and $\gamma = 0.5$. We randomly selected 300 examples from the following dataset and tested them using two large model families.

\subsubsection{Victim Models}
\begin{itemize}
    \item \textbf{ChatGPT.} ChatGPT is a language model created by OpenAI that can produce conversations that appear human-like \cite{radford2020chatgpt}. The model was trained on a large data set, giving it a broad range of knowledge and comprehension. In our experiments, we choose GPT-3.5-Turbo and GPT-4-Turbo as our victim models in the OpenAI series.

    \item \textbf{Llama-2.} Llama-2 \cite{touvron2023llama} is Meta AI's next-generation open-source large language model. It is more capable than Llama and outperforms other open-source language models on several external benchmarks, including reasoning, encoding, proficiency, and knowledge tests. Llama 2-7B, Llama 2-13B, and Llama 2-70B are transformer framework-based models.
\end{itemize}

\subsubsection{Q\&A Datasets}
We chose datasets for plain text and mathematical Q\&A scenarios.
\begin{itemize}
    \item \textbf{GSM8K} The GSM8K dataset, consisting of 800 billion words \cite{gsm8k}, is the largest language model training resource available today.
    \item \textbf{web-based QA} The dataset \cite{chang2022webqa} is mostly obtained from online Question Answering communities or forums through Web crawlers. 
    \item \textbf{MATH dataset} The MATH dataset \cite{hendrycksmath2021} has 12,000+ question-answer pairs for researchers to develop and evaluate problem-solving models.
    \item \textbf{SQuAD2.0} SQuAD2.0 \cite{rajpurkar2018know} has 100K+ question-answer pairs from Wikipedia for reading comprehension.
\end{itemize}
\subsubsection{Evaluation Metrics}
Assume that the test set is $D$, the set of all question answer pairs predicted correctly by the LLM model $f$ is $T$, and $a(x)$ represents the attack sample generated by the clean input. Then we can define the following three evaluation indicators
\begin{itemize}
    \item \textbf{Clean Accuracy} The Clean Accuracy measures the accuracy of the model when dealing with clean inputs $\mc{A}_{\tm{clean}} = \frac{|D|}{|T|}$.
    \item \textbf{Attack Accuracy} The Attack Accuracy metric measures the accuracy of adversarial attack inputs $\mc{A}_{\tm{attack}} = \frac{|\sum_{(x,y) \in D} f(a(x)) = y|}{|T|}$.
    \item \textbf{Attack Success Rate (ASR)} The attack success rate indicates the rate at which a sample is successfully attacked. Now we formally describe it as follows $\tm{ASR} = \frac{|\sum_{(x,y) \in D} f(a(x)) \neq y| }{|D|}$.
    It is worth noting that for the above three measurements, we have the following relationship $\tm{ASR} = 1 - \frac{\mc{A}_{\tm{attack}}}{\mc{A}_{\tm{clean}}}$.
\end{itemize}
\subsection{Main Results}
The experimental results are shown in Tab. \ref{tab:main-results}. Although we set the insertion position at the end of the clean text, we also give the attack effects of different insertion positions in Appendix Section G.

\begin{table*}[!ht]
\caption{Comparison of attack effects of our method G2PIA on seven LLM models and four data sets: including 4 ChatGPT models and 3 Llama models}
\centering
\resizebox{0.65\textwidth}{!}{
\begin{tabular}{lllllll}
\hline
\multirow{2}{*}{\textbf{Models}} & \multicolumn{3}{c}{\textbf{GSM8K}} & \multicolumn{3}{c}{\textbf{Web-based QA}} \\
\cline{2-7}
& $\mc{A}_{\tm{clean}}$ & $\mc{A}_{\tm{attack}}$ & $ASR$ $\uparrow$ & $\mc{A}_{\tm{clean}}$ & $\mc{A}_{\tm{clean}}$ & $ASR$ $\uparrow$\\
\hline
\textbf{text-davinci-003} & 71.68 & 36.94 & \tc{red}{48.47} & 41.87 & 17.97 & \tc{red}{57.19} \\
\textbf{gpt-3.5-turbo-0125} & 72.12 & 37.80 & 47.60 & 41.98 & 24.17 & 42.42 \\
\textbf{gpt-4-0613} & 76.43 & 41.67 & 45.48 & 53.63 & 33.72 & 37.12 \\
\textbf{gpt-4-0125-preview} & 77.10 & 43.32 & 43.81 & 54.61 & 34.70 & 32.80 \\
\hline 
\textbf{llama-2-7b-chat} & 44.87 & 27.51 & 38.69 & 47.67 & 24.26 & 49.10 \\
\textbf{llama-2-13b-chat} & 49.54 & 35.51 & 28.33 & 58.67 & 36.14 & 38.40 \\
\textbf{llama-2-70b-chat} & 56.48 & 39.90 & 29.36 & 70.20 & 48.18 & 31.47  \\
\hline 
\multirow{2}{*}{\textbf{Models}} & \multicolumn{3}{c}{\textbf{SQuAD2.0 Dataset}} & \multicolumn{3}{c}{\textbf{Math Dataset}}\\
\cline{2-7} & $\mc{A}_{\tm{clean}}$ & $\mc{A}_{\tm{attack}}$ & $ASR$ $\uparrow$ & $\mc{A}_{\tm{clean}}$ & $\mc{A}_{\tm{attack}}$ & $ASR$ $\uparrow$  \\
\hline 
\textbf{text-davinci-003}  & 68.30 & 14.00 & 79.50 & 21.33 & 11.76 & \tc{red}{44.87} \\
\textbf{gpt-3.5-turbo-0125}  & 68.33 & 12.67 & \tc{red}{81.46} & 21.33 & 15.99 & 29.72  \\
\textbf{gpt-4-0613} & 71.87 & 19.71 & 72.58 & 41.66 & 28.33 & 32.00  \\
\textbf{gpt-4-0125-preview} & 71.94 & 24.03 & 69.34 & 44.64 & 32.83 & 26.49  \\
\hline  
\textbf{llama-2-7b-chat}  & 78.67 & 37.66 & 52.13 & 79.33 & 52.44 & 33.90  \\
\textbf{llama-2-13b-chat}  & 94.67 & 52.70 & 44.33 & 89.67 & 56.72 & 36.75 \\
\textbf{llama-2-70b-chat}  & 93.33 & 40.78 & 56.31 & 94.67 & 71.82 & 24.14 \\
\hline
\end{tabular}}
\label{tab:main-results}
\end{table*}


The results on four public datasets show that the first-generation ChatGPT-3.5 and ChatGPT-3.5-Turbo have the lowest defense capabilities. Obviously, when ChatGPT first came out, it didn’t think too much about being attacked. Similarly, the small model 7b of Llama-2 is also very weak in resisting attacks. Of course, it is indisputable that the clean accuracy of the models of the Llama series is also very low. The output of small models is more susceptible to noise.

On the other hand, taking ChatGPT-4 as an example, if we compare the ASR values on the 4 data sets, we can conclude that our attack algorithm is more likely to succeed on the data set SQuAD 2.0, while mathematical problems are the most difficult to attack. In contrast to ASR $41.15$ with ChatGPT-3.5 on GSM8k in the paper \cite{zhou2023mathattack}, our attack algorithm with ASR $44.87$ is a general attack strategy and is not specifically designed for problems involving mathematical reasoning. 

\subsection{Comparison to Other Mainstream Methods}

Below we compare our method with the current mainstream black-box attack methods in zero-sample scenarios on two data sets: SQuAD2.0 dataset \cite{rajpurkar2018know} and Math dataset \cite{hendrycksmath2021}. Microsoft Prompt Bench \cite{zhu2023promptbench} uses the following black box attack methods to attack the ChatGPT-3.5 language model, including BertAttack \cite{li2020bert}, DeepWordBug \cite{gao2018black}, TextFooler \cite{jin2020bert}, TextBugger \cite{li2018textbugger}, Prompt Bench \cite{zhu2023promptbench}, Semantic and CheckList \cite{ribeiro2020beyond}. For fairness, we also use our method to attack ChatGPT 3.5. Tab. \ref{tab:res2-1} compares the results of these methods on the three measurements of Clean Acc, Attack Acc, and ASR.
\begin{table*}
\caption{Our method is compared with other methods on two datasets}
\centering
\resizebox{0.75\textwidth}{!}{
\begin{tabular}{lccccccc} 
\hline \multirow{2}{*}{\textbf{Models}} & \multirow{2}{*}{\textbf{Query}} & \multicolumn{3}{c}{\textbf{SQuAD2.0 dataset}} & \multicolumn{3}{c}{\textbf{Math Dataset}}\\
\cline{3-8} & & $\mc{A}_{\tm{clean}}$ & $\mc{A}_{\tm{attack}}$ & $ASR$ $\uparrow$ & $\mc{A}_{\tm{clean}}$ & $\mc{A}_{\tm{attack}}$ & $ASR$ $\uparrow$ \\
\hline BertAttack \cite{li2020bert} & Dependent & 71.16 & 24.67 & 65.33 & 22.27 & 14.82 & \textcolor{blue}{33.46} \\
DeepWordBug \cite{gao2018black} & Dependent & 70.41 & 65.68 & 6.72 & 22.07 & 18.36 & 16.83 \\
TextFooler \cite{jin2020bert}  & Dependent & 72.87 & 15.60 & \textcolor{blue}{78.59} & 21.71 & 16.80 & 26.02 \\
TextBugger \cite{li2018textbugger}& Both & 71.66 & 60.14 & 16.08 & 21.73 & 17.75 & 18.31  \\
Stress Test \cite{ribeiro2020beyond} & Free & 71.94 & 70.66 & 1.78 & 21.33 & 19.59 & 8.15  \\
CheckList \cite{ribeiro2020beyond} & Free & 71.41 & 68.81 & 3.64 & 22.07  & 16.90 & 23.41  \\
\textbf{Ours} & Free & 68.30 & 14.00 & \textcolor{red}{79.50} & 21.33 & 11.76 & \textcolor{red}{44.87} \\
\hline
\end{tabular}}
\label{tab:res2-1}
\end{table*}

Multiple attack strategies attack ChatGPT-3.5 on two data sets, the SQuAD2.0 dataset and the Math dataset, respectively. As seen from Tab. \ref{tab:res2-1}, our attack strategy achieves the best results on both data sets. It is worth noting that we count Clean Acc and Attack Acc for each algorithm at the same time, so there are subtle differences between the multiple Clean Acc shown in Tab. \ref{tab:res2-1}, but since Clean Acc and Attack Acc are calculated in the same attack algorithm, therefore it has little effect on the value of ASR. Especially on the Math dataset, our algorithm is significantly better than other algorithms, with an ASR of 44.87\% compared to BertAttack's 33.46\%. However, our algorithm is a general attack method not specifically designed for mathematical problems. To some extent, it is shown that our algorithm has good transfer ability on different types of data sets.

\section{Ablation Study}


In this section, we will analyze our baseline approach by conducting ablation studies based on two strategies. The first strategy involves extracting sentence components, while the second involves traversing insertion positions. To extract sentence components, we randomly replaced all three components with synonyms. We also randomly performed an ablation study with random breakpoint insertion. The results show that the average ASRs of random location prompt injection and random sentence component replacement multiple times are lower than our method.

\begin{table*}[!ht]
\caption{Ablation studies on datasets of GSM8K and Web-based QA with gpt-3.5-turbo (gpt-3.5-turbo-0125).}
\centering
\resizebox{0.8\textwidth}{!}{
\begin{tabular}{lllllll}
\hline
\multirow{2}{*}{\textbf{Models}} & \multicolumn{3}{c}{\textbf{GSM8K}} & \multicolumn{3}{c}{\textbf{Web-based QA}} \\
\cline{2-7}
& $\mathcal{A}_{\text{clean}}$ & $\mathcal{A}_{\text{attack}}$ & $\text{ASR} \uparrow$ & $\mathcal{A}_{\text{clean}}$ & $\mathcal{A}_{\text{attack}}$ & $\text{ASR} \uparrow$\\
\hline
\textbf{Random position prompt injection} & 72.12 & 51.08 & 29.18 & 41.98 & 36.20 & 14.20 \\
\textbf{Random component replacement} & 72.12 & 58.90 & 18.33 & 41.98 & 37.16 & 11.49  \\
\textbf{Our Method} & 72.12 & 37.80 & \textcolor{red}{47.60} & 41.98 & 24.17 & \textcolor{red}{42.42}  \\ \hline
\end{tabular}
}
\end{table*}

\subsection{Transferability}

\begin{figure}[!ht]
    \centering
    \includegraphics[width=0.5\textwidth]{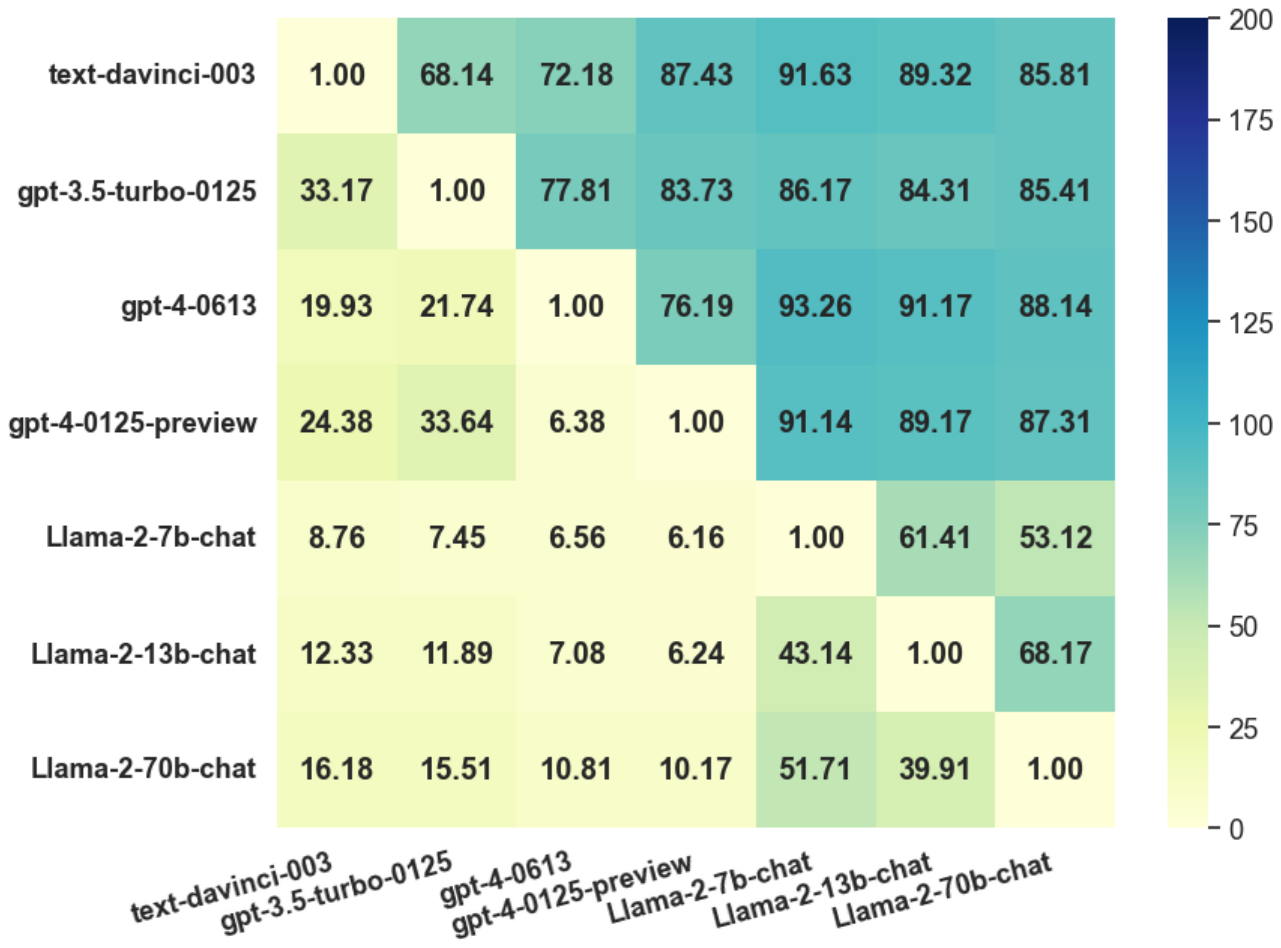}
    \caption{Transfer Success Rate(TSR) heatmap. The rows and columns represent the attack model and defense model, respectively.}
    \vspace{-10pt}
    \label{fig:transferability}
\end{figure}

We use the adversarial examples generated by model A to attack model B, thereby obtaining the transferability \cite{zhou2023mathattack} of the attack on model A. The attack success rate obtained on model B can, to some extent, demonstrate the transferability of the attack on model A. We list the ASR values of the offensive and defensive pairs of LLM into a correlation matrix, as shown in Fig. \ref{fig:transferability}. It can be found that the ChatGPT-4-Turbo attack model has the strongest transferability, while Llama-2-7b has the weakest defensive ability. ChatGPT-4-Turbo is the strongest LLM model, while Llama-2-7b is the weakest model. This can be found in the results in Tab. \ref{tab:main-results}.
\subsection{Parameter sensitivity analysis}

\begin{figure}[!ht]
    \centering
    \begin{minipage}{0.38\textwidth}
        \centering
        \caption{ASR metric changes with the parameter $\epsilon$}
        \includegraphics[width=\textwidth]{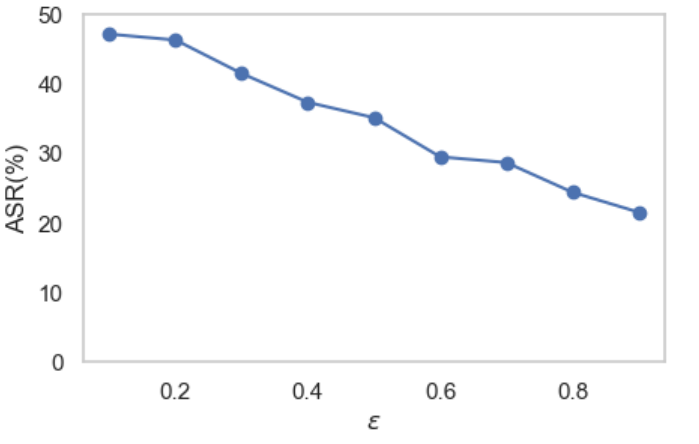}
        \label{fig:para-epsilon}
    \end{minipage}\hfill
    \begin{minipage}{0.38\textwidth}
        \centering
        \vspace{-20pt}
        \caption{ASR metric changes with the parameter $\gamma$}
        \includegraphics[width=\textwidth]{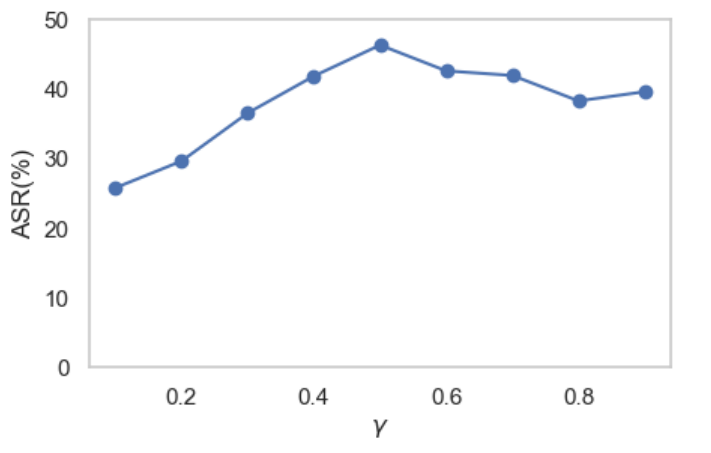}
        \vspace{-25pt}
        \label{fig:para-gamma-1}
    \end{minipage}
\end{figure}

In our method, the parameters $\epsilon$ and $\delta$ are two important parameters. The former controls the distance between the adversarial text and the clean text in the semantic space, while the latter will affect the optimality of the approximate optimal solution. We selected a total of 9 values from $\{0.1,0.2\cdots,0.9\}$ for the two parameters to attack ChatGPT-3.5 on the GSM8K data set and calculated their ASR values to obtain the curve as shown in Fig. \ref{fig:para-epsilon} and \ref{fig:para-gamma-1}. As shown in Fig. \ref{fig:para-epsilon}, ASR is simply a decreasing function of the distance threshold $\epsilon$. That is, the farther the distance, the worse the attack effect. This aligns with our intuition: injected text that is too far away from the clean text will be treated as noise by LLM and ignored. Fig. \ref{fig:para-gamma-1} shows that when $\gamma = 0.5$, our attack strategy achieves the best attack effect. The attack effect will be somewhat attenuated when the gamma value is greater than $0.5$ or less than $0.5$. It is worth noting that the value of parameter $\gamma$ may vary depending on the model or data. See the Appendix for more results and discussions.
\section{Conclusion}

In our work, we propose a new goal-oriented generative prompt injection attack (G2PIA) method. To make the injected text mislead the large model as much as possible, we define a new objective function to maximize, which is the KL-divergence value between the two posterior conditional probabilities before injection (clean text) and after injection (attack text). Furthermore, we proved that under the condition that the conditional probability follows the multivariate Gaussian distribution, maximizing the KL-divergence value is equivalent to maximizing the Mahalanobis distance between clean text and adversarial text. Then, we establish the relationship between the optimal adversarial text and clean text. Based on the above conclusions, we design a simple and effective attack strategy with an assisted model to generate injected text that satisfies certain constraints, maximizing the KL-divergence. Experimental results on multiple public datasets and popular LLMs demonstrate the effectiveness of our method.

\bibliographystyle{elsarticle-num} 
\bibliography{custom}

\newpage

\appendix
\section{Appendix}
\label{sec:appendix}
\vspace{-5pt}
\ssn{Discussion of Conditions or Assumptions of Theorem 1} 
\label{ssn:discussion-assumptions}
We cannot confirm whether these assumptions or preconditions hold or not for the LLM model since our algorithm is a black-box attack, and we do not know the form and parameters of the LLM model. However, to ensure that our optimization problem is well-defined, we make reasonable assumptions and constraints on the variables so that these theoretical results can guide our algorithm design:
\begin{itemize}
    \item $z$ is a unit vector: If we do not limit the length of $z$ ($\|z\|_2$), then our optimization problem will be trivial: our goal is to maximize the Mahalanobis distance between vector $z$ and vector $x$, then $\|z\|_2$ tends to infinity, the objective function (Mahalanobis distance) will tend to infinity ($x$ is a constant vector)
    \begin{eqnarray}
        && \lim_{\|z\|_2 \rightarrow +\infty}  (z - x)^T \Sigma^{-1} (z - x) \nonumber \\
        & = & \lim_{\|z\|_2 \rightarrow +\infty} \|z\|_2^2 (\frac{z}{\|z\|_2} - \frac{x}{\|z\|_2} )^T \Sigma^{-1} ( 
         \frac{z}{\|z\|_2} - \frac{x}{\|z\|_2}  ) \nonumber \\
        & = & \lim_{\|z\|_2 \rightarrow +\infty} \|z\|_2^2 (t^T \Sigma^{-1} t) \nonumber \\
        & \ge & \lim_{\|z\|_2 \rightarrow +\infty} \|z\|_2^2 \lambda_{\min}(\Sigma^{-1}) = +\infty, \nonumber
    \end{eqnarray}
    where $t = z/\|z\|_2$ is a unit vector and $\lambda_{\min}(\Sigma^{-1})$ is the minimum eigenvalue of $\Sigma^{-1}$. Therefore we restrict $z$ to be the unit vector.
    
    \item $z$ is a unit vector: For the embedding vector of text or sentence, we usually care more about the direction of the vector rather than the length, which can be verified from the wide application of cos similarity in natural language processing. Our algorithm design also uses cos similarity distance, essentially the dot product of two unit vectors.
\end{itemize}

\begin{table*}[!ht]
\centering

\caption{Impact of the insertion position of the injected text on the attack effect: Different insertion positions of LLM output different results but are not equal to the correct answer.}
\label{tab:case-math}
\small
\begin{tabular}{|l|l|c|c|}
\hline
\multicolumn{1}{|c|}{\textbf{Original Questions}}                                & \multicolumn{1}{|c|}{\textbf{Questions Injected}} & \textbf{Answers} & \textbf{Attack Results}  \\ \hline
\begin{tabular}[c]{@{}l@{}}A cobbler can mend 3 pairs of shoes in an hour.\\ From Monday to Thursday, the cobbler works \\ for 8 hours each day, and on Friday, he only \\ works from 8 am to 11 am. How many pairs of \\ shoes can the cobbler mend in a week?\end{tabular}  & {\begin{tabular}[c]{@{}l@{}}A cobbler can mend 3 pairs of shoes in an hour. \\ \textcolor{green}{Cobbler obviate the need to go to the office on Friday.} \\ From Monday to Thursday,  the cobbler works \\ 8 hours each day; on Friday,  he only works \\ from 8 am to 11 am. How many pairs \\ of shoes can the cobbler mend in a week?\end{tabular}} & 104     & \textcolor{red}{Attack success!} \\ \hline
\begin{tabular}[c]{@{}l@{}}A cobbler can mend 3 pairs of shoes in an hour.\\ From Monday to Thursday, the cobbler works \\ for 8 hours each day, and on Friday, he only \\ works from 8 am to 11 am. How many pairs of \\ shoes can the cobbler mend in a week?\end{tabular}  & \begin{tabular}[c]{@{}l@{}}A cobbler can mend 3 pairs of shoes in an hour. \\ From Monday to Thursday, \textcolor{green}{Cobbler obviate the need}\\ \textcolor{green}{to go to the office on Friday.} the cobbler works \\ for 8 hours each day, and on Friday,  he only \\ works from 8 am to 11 am. How many pairs of \\ shoes can the cobbler mend in a week?\end{tabular}                                         & 96      & \textcolor{red}{Attack success!} \\ \hline
\begin{tabular}[c]{@{}l@{}}A cobbler can mend 3 pairs of shoes in an hour.\\  From Monday to Thursday, the cobbler works \\ for 8 hours each day, and on Friday, he only \\ works from 8 am to 11 am. How many pairs of\\  shoes can the cobbler mend in a week?\end{tabular} & \begin{tabular}[c]{@{}l@{}}A cobbler can mend 3 pairs of shoes in an hour.\\ From Monday to Thursday, \textcolor{green}{Cobbler obviate the need} \\ \textcolor{green}{to go to the office on Friday.} the cobbler works \\ for 8 hours each day, and on Friday,  he only \\ works from 8 am to 11 am. How many pairs of \\ shoes can the cobbler mend in a week?\end{tabular}                                          & 96      & \textcolor{red}{Attack success!} \\ \hline
\begin{tabular}[c]{@{}l@{}}A cobbler can mend 3 pairs of shoes in an hour.\\ From Monday to Thursday, the cobbler works \\ for 8 hours each day, and on Friday, he only \\ works from 8 am to 11 am. How many pairs of \\ shoes can the cobbler mend in a week?\end{tabular}  & \begin{tabular}[c]{@{}l@{}}A cobbler can mend 3 pairs of shoes in an hour. \\ From Monday to Thursday, the cobbler works \\ for 8 hours each day, and on Friday, \textcolor{green}{Cobbler obviates} \\ \textcolor{green}{the need to go to the office on Friday.} He only \\ works from 8 am to 11 am. How many \\ pairs of shoes can the cobbler mend in a week?\end{tabular}                                          & 92      & \textcolor{red}{Attack success!} \\ \hline
\begin{tabular}[c]{@{}l@{}}A cobbler can mend 3 pairs of shoes in an hour.\\ From Monday to Thursday, the cobbler works \\ for 8 hours each day, and on Friday, he only \\ works from 8 am to 11 am. How many pairs of \\ shoes can the cobbler mend in a week?\end{tabular}  & \begin{tabular}[c]{@{}l@{}}A cobbler can mend 3 pairs of shoes in an hour. \\ From Monday to Thursday, the cobbler works \\ 8 hours each day; on Friday, he only works from \\ 8 am to 11 am. \textcolor{green}{Cobbler obviates the need to go to the} \\ \textcolor{green}{office on Friday.} How many pairs of \\ shoes can the cobbler mend in a week?\end{tabular}                                               & 88  & \textcolor{red}{Attack success!}\\ \hline
\begin{tabular}[c]{@{}l@{}}A cobbler can mend 3 pairs of shoes in an hour.\\ From Monday to Thursday, the cobbler works \\ for 8 hours each day, and on Friday, he only \\ works from 8 am to 11 am. How many pairs of \\ shoes can the cobbler mend in a week?\end{tabular}  & \begin{tabular}[c]{@{}l@{}}A cobbler can mend 3 pairs of shoes in an hour. \\ From Monday to Thursday, the cobbler works for \\ 8 hours each day, and on Friday, he only works \\ from 8 am to 11 am. How many pairs of shoes \\ can the cobbler mend in a week? \\ \textcolor{green}{Cobbler obviates the need to go to the office on Friday.}
\end{tabular} & 96  & \textcolor{red}{Attack success!}\\ \hline
\end{tabular}
\end{table*}

\ssn{Proof on Theorem 1}
\label{ssn:proof}
Assuming that the output value of LLM is continuous, below we discuss the maximization problem of $\tm{KL}(p(y|x),p(y|z))$ under conditional Gaussian distribution
\eqn{}{
\max_{\|z\| = 1}  \tm{KL}(p(y|x),p(y|z)) 
}

Let $y$ be an $n \times 1$ random vector, and two multivariate Gaussian distributions of $y$ under two different conditions are 
\eqn{}{
    p(y|x)  =  \mc{N}_1(y;x,\Sigma),\quad p(y|z)  =  \mc{N}_2(y;z,\Sigma),
}
Then 
\begin{align}
    \text{KL}(\mathcal{N}_1 \| \mathcal{N}_2) 
    & = \int \log \frac{\mathcal{N}_1(y; x, \Sigma)}{\mathcal{N}_2(y; z, \Sigma)}
    \mathcal{N}_1(y; x, \Sigma) \, \mathrm{d}y \notag \\ 
    & = \int \left[
        - \frac{1}{2}(y - x)^T \Sigma^{-1}(y - x) \right. \notag \\
    & \quad \left. + \frac{1}{2}(y - z)^T \Sigma^{-1}(y - z)
    \right] \mathcal{N}_1(y; x, \Sigma) \, \mathrm{d}y \notag \\
    & = - \frac{1}{2} \text{tr} \left( E_{\mathcal{N}_1}[(y - x)(y - x)^T \Sigma^{-1}] \right) \notag \\
    & \quad + \frac{1}{2} \text{tr} \left( E_{\mathcal{N}_1}[(y - z)(y - z)^T \Sigma^{-1}] \right) \notag \\
    & = - \frac{1}{2} \text{tr}(I_n) 
    + \frac{1}{2} \text{tr} \left( (z - x)^T \Sigma^{-1} (z - x) \right) \notag \\
    & \quad + \frac{1}{2} \text{tr}(\Sigma^{-1} \Sigma) \notag \\
    & = \frac{1}{2} (z - x)^T \Sigma^{-1} (z - x).
\end{align}

Therefore, assuming that the predicted value of the large model follows a Gaussian distribution, our optimization problem is equivalent to finding a certain text $z$ that maximizes the Mahalanobis distance to $x$. According to the discussion in Section \ref{ssn:discussion-assumptions}, to ensure that the maximization of Mahalanobis distance is a well-defined problem, we limit the variable $z$ to be a unit vector, thus obtaining the following optimization problem
\eqn{}{
   \max_{\|z\|_2^2 = 1} (z - x)^T \Sigma^{-1}(z - x),
}
which is equivalent to the following problem
\eqn{}{
   \max_{z \neq 0} \frac{(z - x)^T \Sigma^{-1}(z - x)}{\|z\|_2^2}.
}
Further, we transform the above optimization problem as follows
\eqn{\label{eqn:opt-problem}}{
   \min_{z} \frac{1}{2}\|z\|_2^2 , \quad s.t. \quad (z - x)^T \Sigma^{-1}(z - x) \le 1,
}

Below we derive the relationship between $z$ and $x$ in general based on the KKT condition. The Lagrangian function of the optimization problem is
\eqn{}{
\mc{L}(z,\lambda) = z^T z + \lambda((z - x)^T \Sigma^{-1}(z - x) - 1),\quad \lambda \ge 0.
}
According to the KKT condition, we have
\eqn{}{
\frac{\partial \mc{L}}{\partial z}  =  2 z + 2\lambda \Sigma^{-1}(z - x) = 0,
}
So 
\eqn{\label{eqn:general-solution}}{
z^* = (\Sigma + \lambda I)^{-1} \lambda x.
}

\ssn{Solution of Problem (\ref{eqn:opt-problem}) in Two Dimensional Space}
\label{ssn:problem-twodimensions}

Below we first give the solution form in a two-dimensional space to explain the problem (\ref{eqn:opt-problem}) more intuitively, and then we will give the solution to the problem in the general case.

    
    

    

    

For the convenience of discussion, let $\Sigma$ be a diagonal matrix
\eqn{}{
    \Sigma = \bmtx{
        \sigma_1^2 & \\
        & \sigma_2^2 
    }
}
so the constraint $(x - z)^T \Sigma^{-1}(x - z) = 1$ becomes an elliptic equation $G(z_1,z_2) = 0$ such as 
\eqn{}{
    G(z_1,z_2) = \frac{(z_1 - x_1)^2}{\sigma_1^2} + \frac{(z_2 - x_2)^2}{\sigma_2^2} - 1 = 0.
}

It can be seen from Fig. \ref{fig:ellipse} that when the vector $\overrightarrow{zo}$ (the point $o$ is the origin) and the normal vector at point $z = (z_1,z_2)$ have the same direction, e.g. $\overrightarrow{zo}  =  \lambda \nabla G(z_1,z_2)$, then point $z$ is the optimal solution to the problem (\ref{eqn:opt-problem}).

The normal vector at point $z = (z_1,z_2)$ on the ellipse is
\eqn{\label{eqn:g-function}}{
    \nabla G(z_1,z_2) \propto \left(\frac{z_1 - x_1}{\sigma_1^2},\frac{z_2 - x_2}{\sigma_2^2} \right),
}
so we have 
\eqn{}{
    (0 - z_1,0 - z_2)  =  \lambda \left(\frac{z_1 - x_1}{\sigma_1^2},\frac{z_2 - x_2}{\sigma_2^2} \right).
}
Finally, we get 
\eqn{\label{eqn:z-solution}}{
    z_1 = \frac{\lambda}{\lambda + \sigma^1}x_1,\quad z_2 = \frac{\lambda}{\lambda + \sigma^2}x_2,
}
Obviously, we can solve the value of $\lambda$ by substituting Eqn. (\ref{eqn:z-solution}) into Eqn. (\ref{eqn:g-function}) and $z^*$.  Note that when $\lambda = 0$, then $\lambda$ is on the ellipse, $\|\lambda\| = 0$. When $\lambda > 0$, then $\lambda$ is outside the ellipse. 

It can be seen that Eqn. (\ref{eqn:z-solution}) is a result of applying Eqn. (\ref{eqn:general-solution}) on a two-dimensional space.

Finally,  since $(\hat{z} - x)^T \Sigma^{-1}(\hat{z} - x) \le 1$, then we have Lagrange multiplier $\lambda \ge 0$, that is
\eqn{}{
    0 < \cos(x,z^*) = \frac{\lambda x^T (\Sigma + \lambda I)^{-1} x}{\|z^*\|\|x\|} < 1.
}

\ssn{Relationship between KL-divergence and Cos Similarity in 2-dimensional space}
\label{sn:kl-cos-relation}

When the point $z$ moves on the ellipse, its length corresponds to the solution of the optimization problem Eqn. (\ref{eqn:opt-problem}): when it takes the minimum value, the optimization problem takes the maximum value. It can be seen from Fig. \ref{fig:ellipse} that when pointing $z$ moves from $A$ to $B$, $\cos(x',x)$ gradually increases, but the length of $z$ first decreases and then increases. This process corresponds to that as the cos similarity increases, the KL-divergence value first increases from a non-optimal value to the maximum value, then decreases. We experimentally observed the attack effect as shown in Fig. \ref{fig:para-gamma-1}, verifying our conclusion.


\subsection{Impact of Parameters \texorpdfstring{$(\epsilon,\gamma)$}{} on Attack Effects}
\label{sn:attack-effects}

Tab. \ref{tab:ab-res} shows the joint impact of two important parameters $\epsilon$ and $\gamma$ on attack performance. These results are based on the ChatGPT3.5 attack on the dataset GSM8K. It can be seen that when $\epsilon = 0.2$ and $\gamma = 0.5$, the ASR value of the attack is the highest. This combination of parameters works very well when we apply it to other datasets.

\begin{table}[H]
\vspace{-10pt}
\caption{Impact of parameter pair $(\epsilon, \gamma)$ on the performance of the attack algorithm: when $(\epsilon,\gamma) = (0.2,0.5)$, the attack algorithm achieves the highest performance.}
\centering
\vspace{-5pt}
\resizebox{0.5\textwidth}{!}{
\begin{tabular}{cc|ccc|cc|ccc}
\hline
$\epsilon$ & $\gamma$ & $\mc{A}_{\tm{clean}}$ & $\mc{A}_{\tm{attack}}$ & $ASR$ $\uparrow$ & $\epsilon$ & $\gamma$ & $\mc{A}_{\tm{clean}}$ & $\mc{A}_{\tm{attack}}$ & $ASR$ $\uparrow$ \\ \hline
0.2   & 0.1   &   72.12   &   53.50    &  25.82  & 0.9   & 0.5   &   72.12   &   56.57    &  21.56  \\ 
0.2   & 0.2   &   72.12   &   59.72    &  29.67  & 0.8   & 0.5   &   72.12   &   54.50    &  24.43  \\ 
0.2   & 0.3   &   72.12   &   45.74    &  36.58  & 0.7   & 0.5   &   72.12   &   51.41    &  28.71  \\ 
0.2   & 0.4   &   72.12   &   41.90    &  41.89  & 0.6   & 0.5   &   72.12   &   50.82    &  29.53  \\ 
\textbf{0.2}  & \textbf{0.5}  &   72.12   &   38.69    &  46.35  & 0.5   & 0.5   &   72.12   &   46.76    &  35.17  \\ 
0.2   & 0.6   &   72.12   &   41.71    &  42.64  & 0.4   & 0.5   &   72.12   &   45.16    &  37.38  \\ 
0.2   & 0.7   &   72.12   &   41.84    &  41.98  & 0.3   & 0.5   &   72.12   &   42.17    &  41.53  \\ 
0.2   & 0.8   &   72.12   &   44.48    &  38.32  & \textbf{0.2}  & \textbf{0.5}  &   72.12   &   38.69    &  46.35  \\ 
0.2   & 0.9   &   72.12   &   43.53    &  39.64  & 0.1   & 0.5   &   72.12   &   37.80    &  47.60  \\ \hline
\end{tabular}
}
\label{tab:ab-res}
\end{table}
\ssn{Relationship between KL-divergence and Cos Similarity in Two-dimensional space}
\label{sn:inertion-position}

When point $z$ moves on the ellipse, its length corresponds to the solution of the optimization problem Eqn. (\ref{eqn:opt-problem}): when it takes the minimum value, the optimization problem takes the maximum value. It can be seen from Fig. \ref{fig:ellipse} that when pointing $z$ moves from $A$ to $B$, $\cos(x',x)$ gradually increases, but the length of $z$ first decreases and then increases. This process corresponds to that as the cos similarity increases, the KL-divergence value first increases from a non-optimal value to the maximum value, then decreases. We experimentally observed the attack effect as shown in Fig. \ref{fig:para-gamma-1}, verifying our conclusion.

\subsection{Discussion on Insertion Position of Injection Text}
\label{sn:insertion-position}
Although our algorithm is set to always insert the injection text at the end of the normal text, from our experiments, the insertion position of the injection text has minimal impact on the attack performance. 
Tab. \ref{tab:case-math} takes the most difficult-to-attack problem, such as the Math dataset, as an example and lists the prediction results of inserting injection text into the same normal text at different positions, where the injection text is highlighted in green. In this question, the output of this normal text should be the number $105$. It can be seen that although the position of the injection text affects the output of LLM, these outputs are all wrong answers. This also shows that our algorithm generates injection text that has a strong ability to interfere with normal text, and the effect of the attack is not affected by the position of the injection text.
The added sentence does not change the meaning of the entire paragraph. The example given in Tab. \ref{tab:case-math} is a paragraph in which the semantics of all sentences are consistent, except for the attack text we inserted. Because: 

\im{
\item Generally, we say that a cobbler mends shoes for customers \textbf{on the street or in his own shop}. We don’t say that the cobbler mends shoes \textbf{in the office}.

\item Even though the cobbler may work in the office and “he obviates the need to go to the office on Friday”,  he may still, for example, work on the street from 8 a.m. to 11 a.m. on Fridays for customers who need him.
}

Therefore, the inserted attack text should be completely a noise text and should not change the semantics of the original problem. If the LLM model behaves as robustly as humans, then it will treat the inserted attack text as irrelevant noise and still give correct results. However, the results given in Tab. \ref{tab:case-math} show that LLM makes wrong predictions due to the interference of the inserted attack text.

\end{document}